\documentclass[A4paper,11pt]{article}
\usepackage[dvips]{graphicx,epsfig}
\usepackage[round]{natbib}
\usepackage{amsfonts} 
\usepackage{stmaryrd}
\usepackage{fullpage}

\newcommand\traitfin{{\hspace*{1cm}-\hspace*{-2.0mm}-\hspace*{-2.0mm}-\hspace*{-2.0mm}
-\hspace*{-2.0mm}-\hspace*{-2.0mm}-\hspace*{-2.0mm}-\hspace*{-2.0mm}-\hspace*{-2.0mm}-
\hspace*{-2.0mm}-\hspace*{-2.0mm}-\hspace*{-2.0mm}-\hspace*{-2.0mm}-\hspace*{-2.0mm}-
\hspace*{-2.0mm}-\hspace*{-2.0mm}-\hspace*{-2.0mm}-\hspace*{1cm}}}
\begin{document} 

\title{Surface oscillations in channeled snow flows.} 
\author{M. Rastello\footnote{Present address: Laboratoire de M\'ecanique des Fluides et d'Acoustique, UMR CNRS 5509, Ecole Centrale de Lyon/Universit\'e Claude Bernard Lyon 1/INSA Lyon, 
36 avenue Guy de Collongue, 69134 Ecully cedex, France, E-mail:
  marie.rastello@ec-lyon.fr, Fax: +33 (0)4 78 64 71 45, corresponding author}\\ Laboratoire 
de Physique, UMR CNRS 5672,\\ Ecole Normale Sup\'erieure de Lyon, 46 all\'ee d'Italie, 69364 Lyon cedex 07, 
France\\ $\ $\\ A. Bouchet\footnote{Present address: IUT de Marseille,
  D\'epartement Mesures Physiques, 142 Traverse Charles Susini, BP 157,
  13388 Marseille Cedex 13, France, E-mail:
  alexi.bouchet@univ.u-3mrs.fr, Fax.: +33 (0)4 91 28 94 05}\\ Cemagref - UR ETNA,\\ 
2 rue de la papeterie, BP 76, 
38402 Saint Martin d'H\`eres cedex, France}
\date{} 
 
\maketitle 
\abstract{An experimental device has been built to measure velocity profiles and friction laws 
in channeled snow flows. The measurements show that the velocity
depends linearly on the vertical position in the flow and that the friction coefficient is a first-order 
polynomial in velocity (u) and thickness (h) of the flow. In all flows, oscillations on the 
surface of the flow were observed throughout the channel and measured at the location of the probes. The experimental 
results are confronted with a shallow water approach. Using a Saint-Venant 
modeling, we show that the flow is effectively uniform in the
streamwise direction at the measurement 
location. We show that the surface oscillations produced by the Archimedes's screw at the top of 
the channel persist throughout the whole length of the channel and are
the source of the measured oscillations. This last result
provides good validation of the description of such channeled snow
flows by a Saint-Venant modeling.}\\ 
\hspace*{-0.5cm}keywords: snow, channeled flows, surface oscillations,
saint-venant equations

\section{Introduction} 
In the past, many models have been used to describe the behavior of dense snow 
avalanches \citep{voellmy55, perla80, dent83, norem87, savage89, naaim03}. Since snow avalanches are 
very destructive and cannot be controlled, it is very difficult to
fully test these models experimentally. 
Direct measurements on avalanches do exist \citep{dent98} but are rare and must be completed 
by "laboratory experiments" to study the rheology of snow\citep{nishimura89, casassa91, kern04}.
The main problem is that snow grains are very fragile. This brittleness makes it very difficult to use a
large amount of snow in the laboratory. In consequence, as described in 
previous papers \citep{bouchet03a,bouchet03b} we chose to work with an {\it in situ} rheometer, to use
snow directly from the field. To stay as close as
possible to naturally occurring flows in avalanches, we studied free surface snow flows in an inclined channel.

Experiments over a wide range of slopes and thicknesses were performed to measure velocity 
profiles and friction laws in the flow. Results are fully reported in
\citep{bouchet04} and are briefly presented in
section~\ref{profile}. One of the main purpose of the present article is to test the obtained 
snow flows with regards to a commonly used Saint-Venant model to see if the model can help in 
understanding and predicting the behavior of such flows. 

As described later in the article the snow flow
consists of a short first phase ($2$ to $3$ seconds) of transient
regime followed by a fully developed regime which lasts until the end
of the experiment (around $30$ seconds). The fully developed part of
the flow is made on the one hand of a steady component non uniform in
the streamwise direction but which tends to become uniform
asymptotically (x-uniform asymptotic regime) as will be shown in the article. On
the other hand, added to the steady component is a periodic one which
will be studied in detail when looking to the surface oscillations that
are present on top of the flow. The whole experimental and numerical
results presented in the present paper are all focused on the fully
developed regime leaving apart the transient one. The steady component
of the fully developed flow is studied in detail in this article using
the analytical and numerical results. As mentioned before, in all
cases, 
oscillations on the surface of the flow were observed. The question of the origin of 
these oscillations has not been addressed up to now and is another main concern of this paper. Are they 
roll waves intrinsic to the flow like those observed in granular flows
by \citet{louge01}? 
Are they engendered by the feeding system as direct observations suggest? 
Numerical simulations and analytic solutions are used to get a better
understanding of the flow 
characteristics. 

Section~\ref{friction} briefly presents the experimental set-up and
results. The simulation technique is presented in section~\ref{numerique}; it is based on the resolution of the shallow 
water Saint-Venant equations.  The behavior of the flow with
increasing streamwise distance, and in particular the way it
approaches the x-uniform, fully developed regime, is then studied
numerically and analytically. An estimate of the distance needed to
attain the asymptotic regime is derived. Section~\ref{frequences} concerns the 
study of the oscillations on the 
surface of the flow, and the validation of the model.
 
\section{Experimental set-up and results\label{friction}} 
The experimental facility and results are summarized in
\citet{bouchet03b, bouchet04}. A more 
detailed technical description of the set-up, and complete results, can be found in \citet{bouchet03a}. 

\subsection{The experimental device} 
\begin{figure} 
\begin{center} 
\epsfig{file=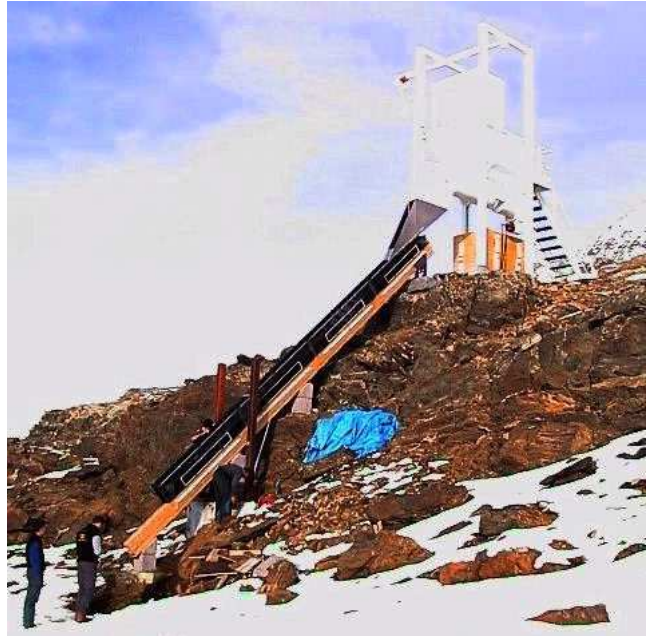,width=8.5cm} 
\caption{The experimental set-up is designed to perform free surface snow flows in a channel (length: 10\,m, width: 20\,cm, thickness:
  20\,cm). The coordinate system associated to describe the flows (see
  figure~\ref{saintvenant}) is the following: the $x$ axis is along the main 
direction of the channel, orientated downward; the $z$ axis is
perpendicular to the bottom of the channel, 
orientated towards the surface of the flow.} 
\label{dispo} 
\end{center} 
\end{figure}

A photograph of the experimental apparatus is shown in figure~\ref{dispo}.
The device consists of a 10\,m long, 20\,cm wide, 20\,cm high channel,
fed with snow by an Archimedes's screw from a hopper at the top of the channel. The screw is driven by a 
thermo-hydraulic engine, so that its rotation speed $f$ is constant
during an experimental run. Such 
a system is able to produce fully developed flows with a typical duration of 30 seconds. The floor of the 
channel is covered with sandpaper to create a rough surface. The slope angle of the channel, 
$\theta$, ranges from 27$^{o}$ to 45$^{o}$, in steps of  2$^{o}$. A
second adjustable parameter is the mean flow thickness ($h$) which is
controlled by the value of the rotation speed f (f ranges from 0\,rpm
to 60\,rpm). Twelve different flows were studied. 
Many types of snow can be found in the field. As a first study the repeatability of the experiments 
was privileged and thus flows were performed in the following conditions. In each case: 
\begin{itemize} 
\item[-] to avoid snow melting the temperature of both snow and air was below 
$-15^{\circ}\mathrm{C}$;
\item[-] snow grains were small and {\em rounded} with diameters between 0.2 and 0.4\,mm;  
\item[-] snow was sieved with a 3\,cm mesh. 
\end{itemize} 
The velocity profile was measured 6\,m downstream from the top, on one
of the sidewalls.
The flow thickness was measured at the same location ($h_{L}$) and also 1.5\,m upstream ($h_{L-1.5}$).

Velocities are deduced~\citep{dent98} from the delay between signals
at two
identical opto-electronic sensors, one 2\,cm downstream from the other.
Each of these sensors consists of a photo-transistor and an LED.
The latter emits an IR light beam which is reflected by the snow and collected by the photo-transistor.
The signal produced by the photo-transistor is thus characteristic of the snow pack flowing in front of the sensor.
Because the distance $d$ between the two sensors is small enough to
avoid deformation of the snow pack, the signals are identical 
but delayed by an amount of time $\tau$. A cross-correlation method
enables measurement of $\tau$, and hence to deduce the velocity $v=d/\tau$.

The flow thickness measuring device (Leuze ODSM/V-5010-600-421) consists of an optical distance sensor positioned above the flow.
It emits an IR light beam (a few degrees) which is reflected by the free surface of the flow and collected by a CCD sensor.
The measured position of the reflected beam on the CCD sensor, is a function of
the distance between the sensor and the surface. Knowing the distance between the sensor and the bottom
of the flow allows the calculation of the flow thickness.

The main characteristics (slope angle, average thickness and frequency of the thickness oscillations) 
of the flows are listed in table~\ref{flows}. 
\begin{table}[h]  
\begin{center}  
\begin{tabular}{|c|c|c|}  
\hline  
slope angle &average value of &F: frequency of the \\  
(degrees) &the thickness $h_L$ (cm) & thickness oscillations (Hz) \\  
\hline \hline  
31 & 8.5 & 0.50 \\  
\hline  
31 & 7.3 & 0.39 \\  
\hline  
33 & 7.4 & 0.44 \\  
\hline  
33 & 8.35 & 0.48 \\  
\hline  
35 & 8.25 & 0.55 \\  
\hline  
35 & 7.6 & 0.50 \\  
\hline  
37 & 8 & 0.68 \\  
\hline  
37 & 9.55 & 0.84 \\  
\hline  
39 & 6.25 & 0.42 \\  
\hline  
39 & 6.05 & 0.38 \\  
\hline  
39 & 6.8 & 0.50 \\  
\hline  
39 & 9.35& 0.68 \\
\hline
\end{tabular}
\caption{Slope angle, average thickness and frequency of the thickness oscillations for each analyzed 
flow.}  
\label{flows} 
\end{center}    
\end{table}  
The frequency $F$ of the oscillations was 
obtained by a Fourier transform of the signal received from the thickness's sensor.   

\subsection{Thickness oscillations, velocity profile and mean velocity\label{profile}} 
\begin{figure} 
\begin{center} 
\epsfig{file=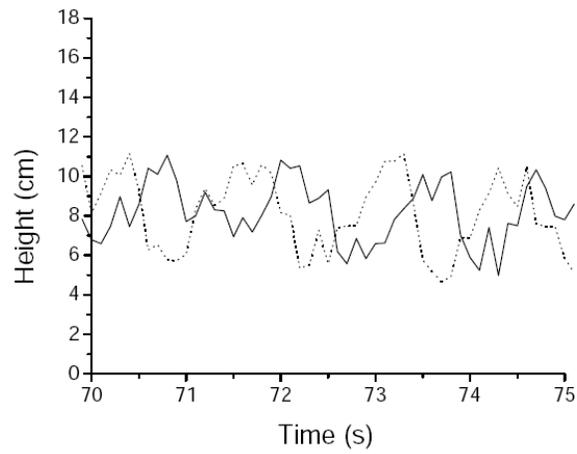,width=8.5cm}
\caption{Example of thickness variations. $h_{L}$ (full line) is measured 1,5\,m downstream from 
$h_{L-1.5}$ (dashed line). The figure corresponds to a laps of 10 seconds taken from a flow 
lasting approximately 30 seconds.} 
\label{height_oscillation} 
\end{center} 
\end{figure}

\begin{figure} 
\begin{center} 
\epsfig{file=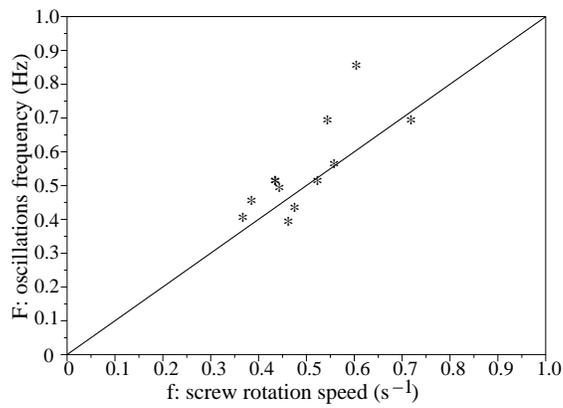,width=8.5cm} 
\caption{Frequency ($F$) of the oscillations observed as a function of the screw rotation speed 
($f$). The straight line is given by $F=f$.} 
\label{Fvsomega} 
\end{center} 
\end{figure}

Initially the channel is empty. The beginning of the experiment 
consists of a transient phase, while the front of the flow goes
throughout the channel. After this transient phase, as shown in
figure~\ref{height_oscillation}, 
at any given location the flow thickness oscillates in time around a mean value $<h_x>$ defined
as a time average (steady component of the flow). 
As also illustrated in
figure~\ref{height_oscillation}, the variations of $h_L$ follows those
of $h_{l-1.5}$ with a constant delay $\tau$: $h_{L}(t) \approx
h_{L-1.5}(t-\tau)$. It thus appears that the steady component of the
flow is x-uniform at the location of the probes. The frequency of the thickness
oscillations mentioned above is always close to the rotation speed of
the screw, as illustrated by figure~\ref{Fvsomega}. 
Indeed, the
rotation speed of the screw can be calculated according to the
formula: $f= {\cal V}^{-1}\times Q$ with $Q$ the flow rate (obtained
from the velocity profile as described below) and ${\cal
  V} \approx 0.14$~m$^{3}$ (information given by the manufacturer). $f$ is 
calculated with a $10\%$ error.

Special attention was payed to side wall effects on the flow. First the existence of a transverse 
velocity gradient was qualitatively looked for. To that purpose the flow was filmed from above. The observation show a 
displacement of the snow as a block without slowdown toward the walls. More quantitatively \citet{louge01} measured 
velocity gradients in a granular channeled flow and obtained less than $10\%$ in velocity between the center of the flow 
and the side walls. Because in the present case the material is cohesive the gradient should be even less than these 
$10\%$. Side walls could also be responsible for a global slowdown of the flow. \citet{ericksson55} performed measurments 
of friction coefficients between snow and different materials. The friction coefficient between snow and plexiglas 
was measured to be around $0.05$ which is ten times lower than the friction between snow and the sandpaper sticked on the 
bottom of the channel. Thus, it can be assumed that the effect on the mean velocity is negligible. From these two 
points, the flow can be considered as uniform in the direction perpendicular to the flow.

The
mean velocity profile is written $V(z)$, where $z$ is the distance
from the bottom of the channel. The results for the twelve different flows are all satisfyingly fitted by a linear 
profile \citep{bouchet04}: 
 
\begin{equation} 
V(z;\theta , h) =V_{s} + \Gamma_{0} z 
\label{eq_profil} 
\end{equation} 
where:
\begin{equation} 
V_{s} \left( \theta\right)=V_{0}\left[ \tan \theta - \tan \theta_{0}  \right]  
\label{eq_slip} 
\end{equation} 
and
\begin{eqnarray} 
\left\{ \begin{array}{lcl} 
V_{0} & = & 9.1 \textrm{ m.s$^{-1}$} \\ 
\tan{\theta_{0}} & = & 0.28 \textrm{ } (\theta_{0}=15.6^{o}) \\ 
\Gamma_{0} & = & 12.5 \textrm{ s$^{-1}$} 
\end{array} \right. 
\label{eq_param} 
\end{eqnarray} 
Such a linear profile is not usual but has already been observed in some granular 
flows~\citep{douady02,bonamy02a}. \citet{bonamy02b} explained this uniformity of the velocity gradient by the 
presence, in the flow, of aggregates of many different sizes without one typical scale. In the present flow, aggregates were also detected~\citep{bouchet03a}. It should be mentioned that the behavior of the flow 
near the ground was not explored. Indeed, the first transducer is located more than ten to twenty particles diameters 
above the ground. Thus, it is here impossible to know if there is a basal sliding or a strong basal shear. The presence 
of the sandpaper on the bottom makes the second hypothesis the more believable. Nevertheless in the present article 
we are only interested in the averaged velocity over the flow thickness which is quite insensitive to this basal 
behavior. Moreover, during the numerical calculation the shape factor is taken equal to one (see 
section~\ref{numerique}).

The integral of equation~(\ref{eq_profil}) gives the averaged
velocity over the flow thickness: 
\begin{equation} 
u \left( \theta,h \right) = V_{s} \left( \theta\right) + \frac{1}{2}\Gamma_0 h 
\label{eq_Uavg1} 
\end{equation} 
or, using equation~\ref{eq_slip},  
\begin{equation} 
u = V_{0}  \left[ \tan \theta - \tan \theta_{0} \right] + \frac{1}{2}  
\Gamma_{0}  h .
\label{eq_Uavg2} 
\end{equation}

\subsection{Friction law\label{friction_sub}} 
Because the mean flow near the probes is fully developed and x-uniform, the friction force 
exerted by 
the channel on the flow balances the weight of the snow. The effective friction coefficient 
$\mu$, defined as the ratio between the shear stress and the normal stress at the bottom of the 
channel, is thus equal to the tangent of the slope angle $\theta$.
Using equation~(\ref{eq_Uavg2}), the friction law can be written as: 
\begin{equation} 
\mu = \mu_{0} + \frac{u}{V_{0}} - \frac{h}{H_{0}} 
\label{eq_friction} 
\end{equation} 
with
\begin{eqnarray} 
\left\{ \begin{array}{lclcl} 
\mu_{0} & = & \tan \theta_{0} & = & 0.28 \\  
H_{0} & = & 2 V_{0} / \Gamma_{0} & = & 1.45\,\mathrm{m} 
\end{array} \right. 
\end{eqnarray} 
It must borne in mind that this friction law is here established for the particular dynamical 
conditions of a fully developed and x-uniform flow. Another point is that this 
friction law has been established in a given range of slope angles and flow thicknesses 
corresponding to those of the 
experiments and should not be extrapolated to some others too far as for example too large values of the thickness. The 
non-transient phase of the following simulations is performed close to this range of validity.
The initial transient phase corresponds to small flow thicknesses. Thus, the friction law is
supposed valid even during this phase in our simulations. 

\section{Study of the steady component of the flow\label{numerique}} 
In this section we study in detail the behavior of the steady part of
the fully developed flow. To that purpose we use the laterally unconfined 
Saint-Venant
equations solving them first numerically and in a second time
analytically.

\subsection{The Saint-Venant equations} 
Originally used for hydraulic studies of shallow water flows, the Saint-Venant equations were first 
applied to granular flows by \citet{savage89}. 
\begin{figure} 
\begin{center} 
\epsfig{file=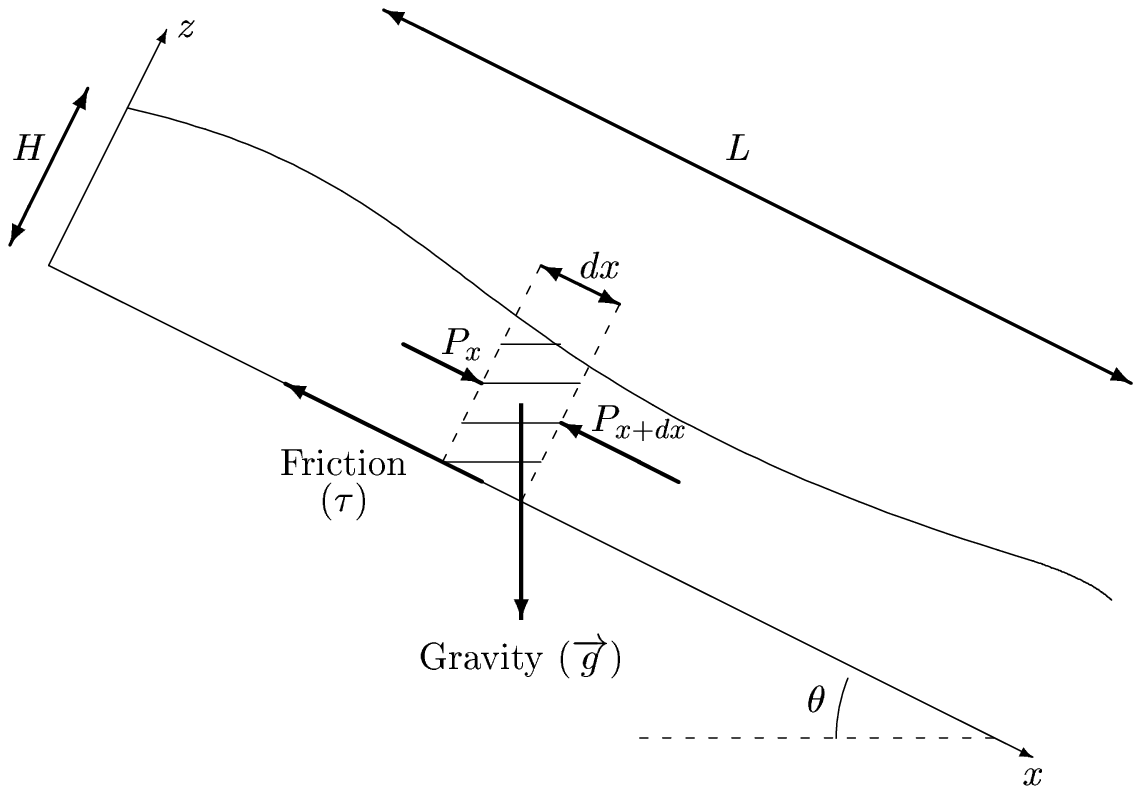,width=8.5cm}
\caption{Sketch of the flow and of the applied forces.} 
\label{saintvenant} 
\end{center} 
\end{figure}
This approach is now commonly 
employed to model dense snow flows. The equations are derived by
averaging the mass and 
momentum conservation equations over the 
flow thickness. Two assumptions are employed: 
\begin{itemize} 
\item the typical flow thickness is assumed to be small compared to
  the streamwise variations (short wavelength hypothesis); 
\item the flow density is taken as constant and uniform. 
\end{itemize} 
Hence, the dominant component of flow velocity, u, is parallel to the bottom
of the channel. The flow is described by its thickness $h(x,t)$ and average 
velocity \mbox{$u(x,t) = (\int_{0}^{h} u(x,z,t) dz )/h$}. Mass and
momentum equations for a thin slice can be written: 
 
\begin{eqnarray} 
\frac{\partial h}{\partial t} + \frac{\partial h  u }{\partial x} & = & 0 \label{masse.finale} \\ 
\frac{\partial h   u }{\partial t} + \alpha\frac{\partial h 
  u^{2} }{\partial x} & = & 
g  h  \cos \theta  (\tan \theta - \mu -k \frac{\partial h}{\partial x}) \label{qmd.finale} 
\end{eqnarray} 
 where
\begin{itemize} 
\item $\alpha$ is the shape factor: $\alpha = h  \int_{0}^{h} u^{2} dz / \left[ \int_{0}^{h} u 
dz \right]^{2}$; 
\item $k$ is the ratio between the normal stress parallel to the ground, $\sigma_{xx}$, and the 
normal stress perpendicular to the ground, $\sigma_{zz}$.
\end{itemize} 
 
In what follows, the stress tensor is supposed to be isotropic and thus $k$ is taken equal to one. In granular flows, 
the precise value of $k$ does not appreciably affect the flow
behavior. For commonly encountered velocity profiles (uniform, linear,
parabolic), $\alpha$ takes values between 1 and 2. The flow is also
insensitive to the precise value used for $\alpha$; here we take
$\alpha =1$. Finally friction law~(\ref{eq_friction}) is used.

\subsection{Description of the numerical solution procedure and result} 
The 6-meter long channel upstream of the measuring device is
discretized in $x$ with points every five 
centimeters. The time step is $dt=5\times 10^{-3}~\mathrm{s}$, and we run 
the model for thirty seconds. The numerical scheme used is an explicit 
finite difference MacCormack scheme.  

Both steady-inlet and initial conditions are needed for $h$ and $u$. 
As regards the steady-inlet conditions ($h(0)$, $u(0)$), once the flow
becomes steady equation~(\ref{masse.finale}) implies a constant volume flux:
\begin{equation} 
h(0)u(0)=h(L,t)u(L,t)=Q\label{debit.eq} 
\end{equation}
Experimental values of $h_L$ and $u_L$ (given by (\ref{eq_Uavg2})) allow
$Q$ to be determined. $h(0)$ is taken as an input parameter for the
simulation and $u(0)$ calculated using $u(0)=Q/h(0)$. These constant
values of $h(0)$ and $u(0)$, which apply strictly to the steady
component of the flow, 
are imposed throughout the calculation. As can be observed in
section~(\ref{uniforme_sec}) the value used for $h(0)$ has no influence
on the asymptotic flow of the end of the channel. 

\begin{figure} 
\begin{center} 
\epsfig{file=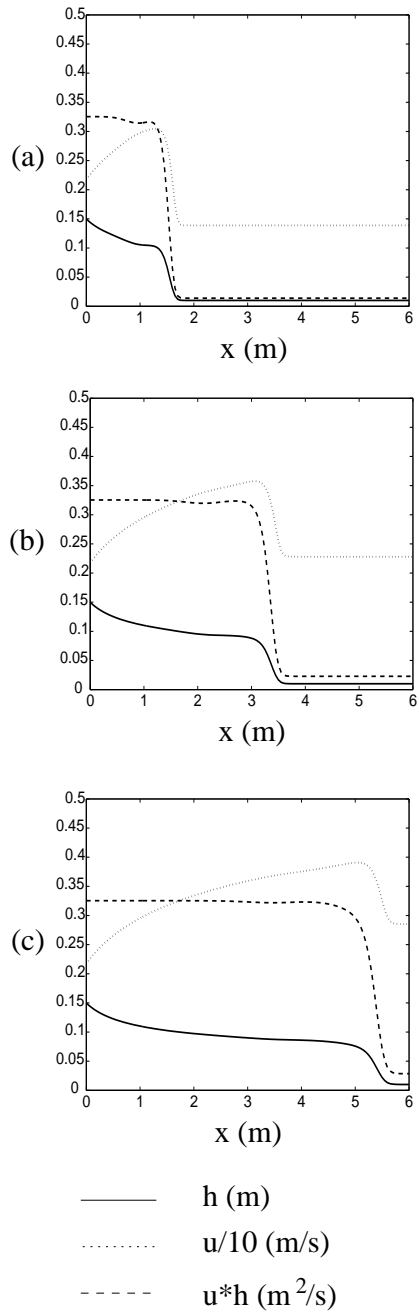,width=6cm}
\caption{Saint-Venant model's results using $h_0=15\,\mathrm{cm}$. thickness $h$ (in m), velocity $u/10$ 
(in m/s) and flow rate  $uh$ (in m$^{2}$/s) are 
plotted versus position $x$. These results concern the beginning of the flow 
when the front is crossing the channel before the establishment of the fully developed regime. (a): 
$t=0.5~\mathrm{s}$, (b): $t=1~\mathrm{s}$, (c): $t=1.5~\mathrm{s}$.
\label{obtention_stationnaire}}  
\end{center} 
\end{figure}

\begin{figure} 
\begin{center} 
\epsfig{file=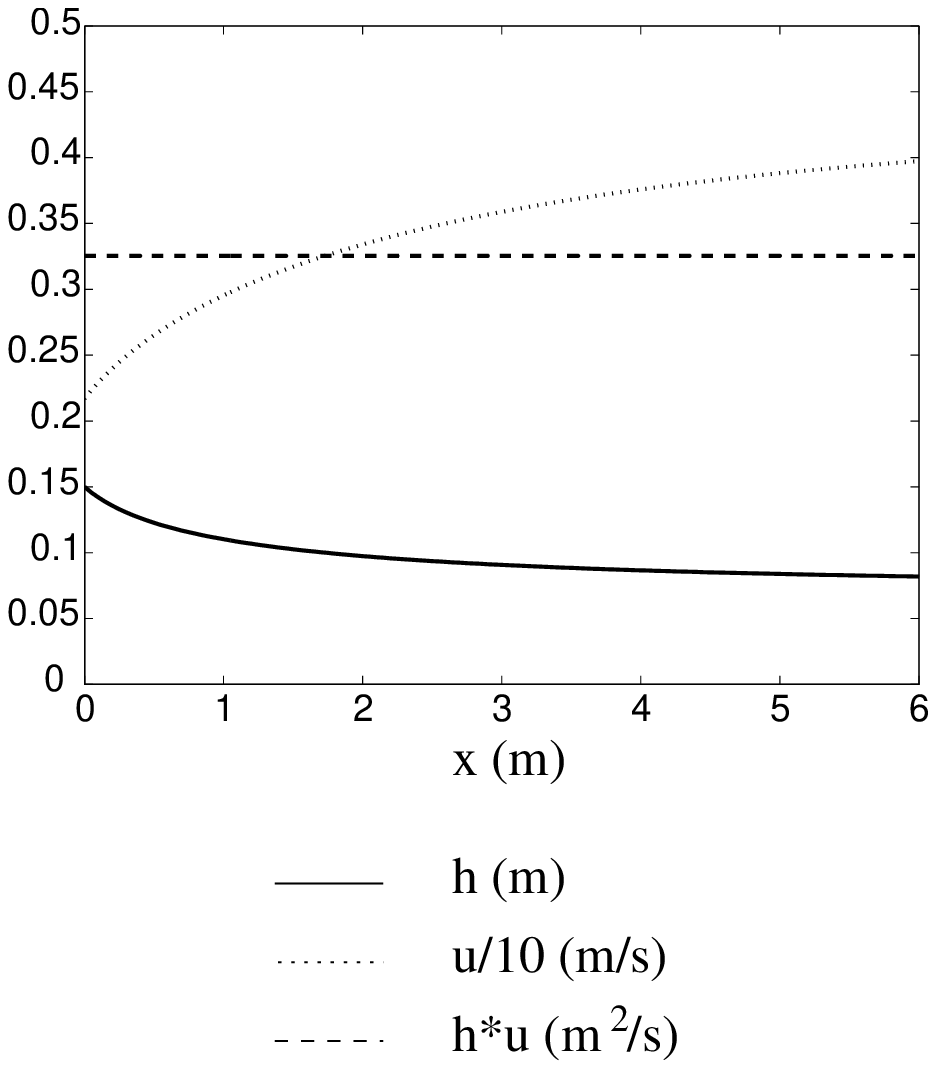,width=8.5cm} 
\caption{Saint-Venant model's results using $h_0=15\,\mathrm{cm}$, thickness $h$ (in m), velocity $u/10$ 
(in m/s) and flow rate $uh$ (in m$^{2}$/s) are 
plotted versus position $x$(m). These results are obtained in the fully developed regime.
\label{stationnaire_fig}}
\end{center} 
\end{figure}

The values used for $\theta$ and $h_L$ are the mean of those encountered during the 
experiments: 
\begin{equation}
h_L=7.6\,\mathrm{cm}
\end{equation}
and 
\begin{equation}
\theta =35^{\circ}
\end{equation} 
When the model is run, an initial unsteady, non-uniform phase is
observed, corresponding to the passage of the front along the channel
(see figure~\ref{obtention_stationnaire}). 
This phase is not discussed
because the Saint-Venant equations cannot deal well with
rapid changes in thickness and velocity. The only thing that can be
noticed is that the passage of the transient phase lasts around $2$ to
$3$ seconds both in the experiments and in the numerics. Once the front has passed, a
fully developed steady regime is observed (see for example figure~\ref{stationnaire_fig}).    
Near the end of the channel the flow is found to be x-uniform. This is
an important point for the exploitation of the experimental results in
general and of the friction law in particular. This
regime is examined in more detail in
the following section. Just for information, the initial values are
$h(x>0,t=0)=1~\mathrm{cm}$ and $u(x>0,t)=0$. The non-zero value of $h$
was introduced to avoid a division by zero. We choose these values for
the sake of simplicity and because the fully developed
flow (the one which is relevant to study here) is not sensitive to them.  

\subsection{Analytical study of the steady component of the
  flow\label{uniforme_sec}}
In this section, we solve analytically the Saint-Venant equation for
the steady component of the flow. This enables us to quantify when the
asymptotic x-uniform regime is reached. 
For steady flows, the Saint-Venant equations (equations~(\ref{masse.finale}) and 
(\ref{qmd.finale})) take the form: 
 
\begin{eqnarray} 
u  h & = & Q \label{masse.stat} \\ 
\alpha\frac{d h  u^{2} }{d x} & = & g   h  \cos \theta (\tan \theta - \mu_{0} - \frac{u}{V_{0}} 
+ \frac{h}{H_{0}} - k \frac{d h}{d x}) \label{qmd.stat} 
\end{eqnarray} 
 
\subsubsection{Values of the x-uniform asymptotic regime} 
 
Let $h_{\infty}$ and $u_{\infty}$ be the flow thickness and flow
velocity when the x-uniform, asymptotic regime is reached. From equations~(\ref{masse.stat}) and (\ref{qmd.stat}), 
$h_{\infty}$ and $u_{\infty}$ are given by: 
 
\begin{eqnarray} 
\left\{ \begin{array}{l} 
h_{\infty}u_{\infty} = Q \\ 
\displaystyle \mu_{0} + \frac{u_{\infty}}{V_{0}} - \frac{h_{\infty}}{H_{0}} = \tan \theta 
\end{array} \right. 
\end{eqnarray} 
 
This system admits only one set of positive solutions: 
 
\begin{eqnarray} 
\left\{ \begin{array}{lcl} 
h_{\infty} & = & \displaystyle \frac{H_{0}  (\tan \theta - \mu_{0})}{2}  \left( -1+\sqrt{1+
\varepsilon} \right)  \\ 
\\ 
u_{\infty} & = & \displaystyle \frac{V_{0}   (\tan \theta - \mu_{0})}{2} \left( 1+\sqrt{1+
\varepsilon} \right) 
\end{array} \right. 
\end{eqnarray} 

where 
 
\begin{equation} 
\varepsilon = \frac{4  Q}{V_{0}  H_{0} (\tan \theta - \mu_{0})^{2}}
\end{equation} 
 
\subsubsection{Distance needed to reach the x-uniform asymptotic regime} 
\begin{figure} 
\begin{center} 
\epsfig{file=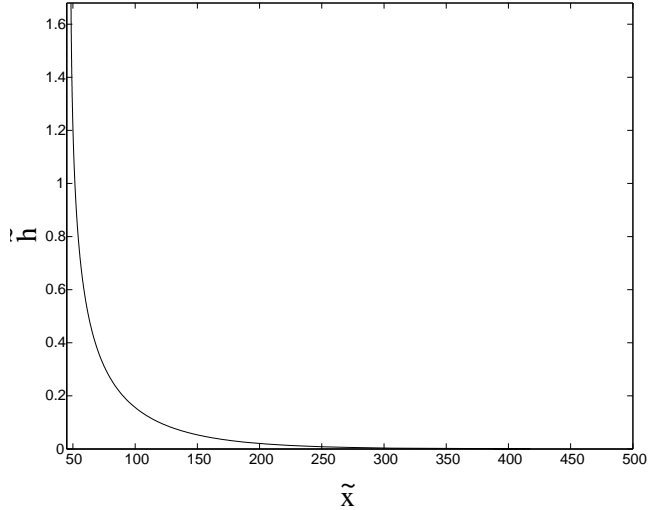,width=8.5cm} 
\caption{Plot of the relative deviation of the flow thickness $\tilde{h}$ as a function of 
the relative distance $\tilde{x}$.\label{w_fig}} 
\end{center} 
\end{figure}

We next study the development of the flow with streamwise distance
$x$. To that purpose we
integrate the differential equations~(\ref{masse.stat})
and~(\ref{qmd.stat}) analytically using the following reduced variables:
$\tilde{h}=(h-h_{\infty})/h_{\infty}$,
$\tilde{u}=(u-u_{\infty})/u_{\infty}$, with
$\tilde{x}=(x+x_0)/h_{\infty}$ with $x_0$ a constant value. Details can be found in
appendix 1. The integration leads to 
\begin{eqnarray}
\lefteqn{\mbox{\hspace*{-1cm}}\ln \left[ \left| \tilde{h}\right|^{k-\alpha  Fr^{2}}  \left| \tilde{h}+1 \right|^{\alpha  Fr^{2}  
\left( 1-1/\beta^{2} \right) }  \left| \tilde{h}+1+\beta \right|^{k   \beta + \alpha  Fr^{2}/\beta^{2}} 
\right]}\nonumber \\
&&-\alpha  Fr^{2}  \frac{1+\beta}{\beta}  \frac{1}{\tilde{h}+1}
= \left( 1 + \beta \right)  \frac{h_{\infty}}{H_{0}}  \tilde{x}
\label{horreur} 
\end{eqnarray} 
 
with 
 
\begin{eqnarray} 
\left\{ \begin{array}{lcl} 
h_{\infty} & = & \displaystyle \frac{H_{0}  \left( \tan \theta - \mu_{0} \right)}{2}  \left[ -1 
+ \sqrt{1+\varepsilon} \right] \\ 
u_{\infty} & = & \displaystyle \frac{V_{0}  \left( \tan \theta - \mu_{0} \right)}{2}  \left[ 1 
+ \sqrt{1+\varepsilon} \right] \\ 
\varepsilon & = & \displaystyle \frac{4Q}{V_{0}  H_{0}  \left( \tan \theta - \mu_{0} 
\right)^{2}}\\ 
Fr^{2} & = & \displaystyle \frac{u_{\infty}^{2}}{g  h_{\infty}  \cos \theta}\\ 
\beta & = & \displaystyle \frac{u_{\infty} 
  H_{0}}{h_{\infty}  V_{0}} = \frac{\left( 1+\sqrt{ 1
+\varepsilon} \right)^{2}}{\varepsilon} 
\end{array} \right. 
\label{horreur_param} 
\end{eqnarray} 
and 
\begin{equation}
x_0=\frac{H_0}{1+\beta}\times (\ln [| \tilde{h}(0)|^{k-\alpha  Fr^{2}}  | \tilde{h}(0)+1 |^{\alpha  Fr^{2}  
( 1-1/\beta^{2} ) }  | \tilde{h}(0)+1+\beta |^{k   \beta + \alpha  Fr^{2}/\beta^{2}} 
]-\frac{\alpha  Fr^{2}  (1+\beta)}{\beta(\tilde{h}(0)+1)}
)
\end{equation}
Using equation~(\ref{horreur}) we plot $\tilde{h}$ as a function of
$\tilde{x}$ in figure~\ref{w_fig}. 
The entire curve has three
branches. Because the thickness of the flow is limited by the vertical
size of the channel ($h<20~\mathrm{cm}$) and because the flow
thickness decreases experimentally throughout the flow we only keep
the part of the curve which fits these conditions. Looking at the curve, one can 
see that when $\tilde{x}$ becomes large enough, the flow reaches asymptotically its x-uniform 
state. This confirms what the numerics seem to predict. The question is now to know whether
the asymptotic x-uniform regime is reached before the end of the
channel. To that purpose we simplify equation~(\ref{horreur}) for the
case of large $\tilde{x}$. It can be rewritten in:
\begin{equation}
\tilde{h}=(1+\beta)^{-(k\beta+\frac{\alpha
    Fr^2}{\beta})}exp\left(-\frac{(1+\beta)h_\infty}{H_0 (\alpha Fr^2-k)}\tilde{x}\right)
\end{equation}
The typical length scale is then:
\begin{equation}
L=\frac{H_0(\alpha Fr^2-k)}{(1+\beta)}=4~\mathrm{m}
\end{equation}
when taking the experimental value $Q=0.32~\mathrm{m^2.s^{-1}}$. It is
therefore relevant to consider the flow as x-uniform at the location of the measuring 
device, 6\,m downstream from the top of the channel.  
 
\section{Origin of the oscillations - validation of the model\label{frequences}} 
In this section we show that the model gives an explanation about the origin of the oscillations observed in the 
experiments.
 
\subsection{Frequency of the oscillations} 
\begin{figure} 
\begin{center} 
\epsfig{file=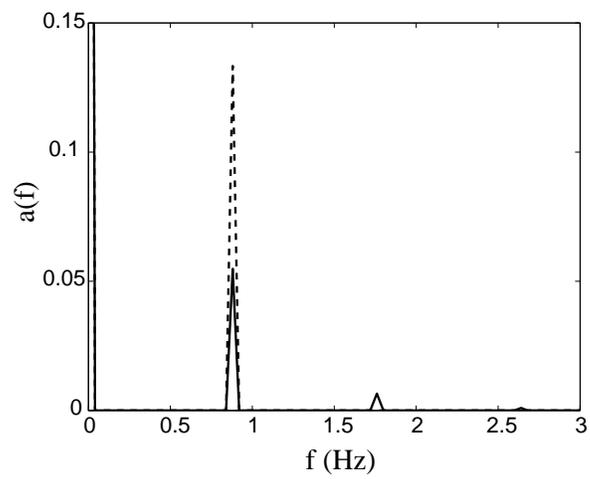,width=8.5cm}
\caption{Fourier-transform amplitude of the thickness of the flow obtained by the Saint-Venant model. 
$- - -$ Fourier-transform amplitude of $h(0,t)$, \traitfin
Fourier-transform amplitude of $h(L,t)$. 
\label{attenuation_fig}}  
\end{center} 
\end{figure}

The oscillations induced by the screw which feeds snow into the
channel are modeled by the boundary condition $h(0,t)=h_0+h_1 \cos
(\omega t)$ with $f=\omega /2\pi$ the frequency of the screw. 
To study what 
happens at the end of the channel we use the  following scheme: 
\begin{itemize} 
\item First, the model is run for 5 seconds with $h(0,t)=h_0+h_1 \cos
  \omega t$ to allow time for the initial transient phase to have finished. 
\item After these 5 seconds the model is run for a further 30 seconds
  and the temporal Fourier transform of $h(L,t)$ is taken.
\end{itemize} 
An example of the results is shown in figure~\ref{attenuation_fig},
which also gives the result for $h(0,t)$. 
In addition to the 
fundamental frequency f and the zero frequency peak representing the
mean value, the harmonics $2f$, $3f$, ... are also present in $h(L,t)$
due to nonlinearity of equations~(\ref{masse.finale}) and
~(\ref{qmd.finale}). The amplitude of these frequency components is always 
less than $10\%$ of the amplitude of the  fundamental peak.  
 
\subsection{Comparison between model and experiments} 
\begin{figure} 
\begin{center} 
\epsfig{file=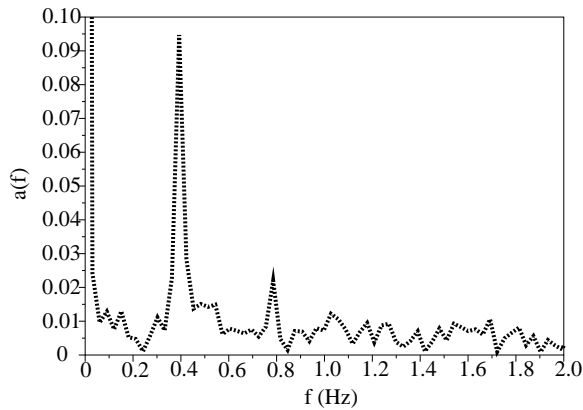,width=8.5cm}
\caption{Fourier-transform amplitude of $h(L,t)$ with $h(L,t)$ measured during one experiment.
\label{attenuation_exp}} 
\end{center} 
\end{figure}

The Fourier transform of $h(L,t)$ in the experiments also shows the 
presence of the frequencies $2f$, $3f$... (cf figure~(\ref{attenuation_exp})).  
However, the ratio of the amplitude of these secondary peaks to that
of the fundamental differs in the model and in the experiments. In both cases (numerical 
and experimental) the secondary peaks are weak and can thus be
neglected. 

\begin{figure} 
\begin{center} 
\epsfig{file=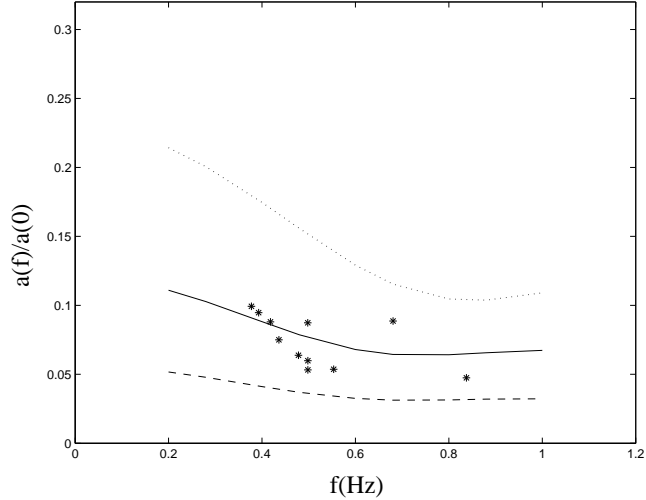,width=8.5cm}
\caption{Ratio between the amplitude of the fundamental peak and the
  amplitude of the peak at zero frequency versus 
frequency. $*$ results from experiments, \traitfin results of the Saint-Venant model 
with $h_0=15~\mathrm{cm}$ and $h_1=4~\mathrm{cm}$, $\cdots\cdots$ results of the Saint-Venant 
model with $h_0=12~\mathrm{cm}$ and $h_1=6~\mathrm{cm}$, $- - -$ results of the Saint-Venant 
model with $h_0=16~\mathrm{cm}$ and $h_1=2~\mathrm{cm}$. \label{comparaison_fig}} 
\end{center} 
\end{figure}

Let us now compare the ratio of 
the amplitude of the fundamental peak to that at zero frequency.  
This ratio is plotted in figure~\ref{comparaison_fig}. 
With $h_0=15~\mathrm{cm}$ and 
$h_1=4~\mathrm{cm}$ the model gives good results compared with the experimental values. The curves 
given by the extreme values of the parameters are also plotted on figure~\ref{comparaison_fig} 
($h_0=12~\mathrm{cm}$, $h_1=6~\mathrm{cm}$ (eg) $h_0/h_1=2$ and $h_0=16~\mathrm{cm}$, 
$h_1=2~\mathrm{cm}$ (eg) $h_0/h_1=8$). By ``extreme values'' we mean
the larger or smaller values of the parameters the experimental
observations allow to use. When plotting the curves what appears to be important is 
not the values of $h_0$ and $h_1$ separately but the ratio $h_0/h_1$. Indeed, with the same 
ratio, two sets of values ($h_{0a}$, $h_{1a}$), ($h_{0b}$, $h_{1b}$) give merely the same 
curves. This is the reason why we have plotted only two ``extreme'' cases. One can see that if 
the experimental values are close to the predictive curve they are also all included in the 
space between the two extreme curves. This allows for different conclusions. The majority of 
points (those that are close to the predictive curve) show that the model well predicts the 
behavior of the flow. The model is run in each case with fixed parameters ($\theta =35^{\circ}$, 
$h_{L}=7.6~\mathrm{cm}$). These values are the mean of the 11 first experimental values 
(see table~\ref{flows}), the twelfth amplitude was not usable. Hence it seems reasonable that 
some of the points are not strictly close to the predictive curve. Nevertheless they are still 
between the extreme curves. All of this confirm the hypothesis that the model well predicts the 
phenomenon and the flow and that the oscillations are due to the screw. 
  
\section{Conclusion} 

The dense snow flows studied during the experiments exhibit thickness's oscillations. Though 
experimental facts suggest that they are generated by the feeding system of the set-up, the 
origin of these oscillations remain {\em a priori} uncertain. The friction law for dense snow flows 
deduced from the experiments is used in numerical resolutions of Saint-Venant's equations. The 
initial conditions given to the thickness at the entrance of the channel consist in sinusoidal 
oscillations with an amplitude $h_{1}$, around a mean value $h_{0}$. The simulations performed 
lead to the following results:
\begin{itemize}
\item the mean flow is x-uniform at the location of the measuring
  device. This is a crucial point for the validation of the results of
  the experiments in which the hypothesis of x-uniformity was the
  central hypothesis;
\item the oscillations engendered at the channel's entrance propagate and are still present with 
the same frequency (harmonics can be neglected) at the measurements' position; 
\item numerical and experimental results are in a good agreement when using an oscillation ratio 
$h_{1}/h_{0}$ = 4/15. From direct observations, this ratio seems very reasonable;
\item the relative intensity of these oscillations (given by the FFT of $h(x,t)$) reduces from 
about 15 $\%$ at the top of the channel to approximatively 5 $\%$ to 10 $\%$ at the studied 
position.
\end{itemize}
The results presented in this paper agree with the experimental observations. They suggest that:
\begin{enumerate}
\item in the experiments, the mean value of the thickness of the flow at the entrance of the
  channel can be changed as wanted
  without affecting the x-uniformity of the flow at the end of the
  channel and thus the validity of the measurements; 
\item reasonable oscillations generated by the feeding system can propagate all along the 
channel and have the same intensity as the one observed in the experimental results;
\end{enumerate}
The present study reveals that, as many other flows, such channeled flows of 
dry snow can be described using a Saint-Venant like model. Such numerical studies can help 
in the future in preparing other configurations of the device to test other caracteristics 
of those channeled snow flows and thus rheology of snow.
$\ $\\

\thanks{A. Bouchet thanks the Cemagref for providing a major financial support to his 
phd-thesis. The authors thank the ETNA research unit for supplying the experimental device. 
Special thanks are due to Mohamed Naaim for the idea of the experiment and for the money to 
realize it. This work was partially supported by an ``ACI Risques
Naturels'' grant and by two grants of the ``P\^ole Grenoblois sur les
Risques Naturels''. Jean-Louis Mari\'e is thanked for his advices
during the improvement of the manuscript, and many thanks are due to
Professor Julian Scott for the improvement of the english wording and of the
article in general.}

\newpage
\section*{Appendix 1}
Using the mass conservation equation 
(equation~(\ref{masse.stat})), equation~(\ref{qmd.stat}) can be rewritten in:

\begin{equation} 
\frac {d}{dx} \left[ \alpha   Q   u + \frac{k   g 
    \cos \theta}{2}   h^{2} \right] = g  h  \cos \theta  
\left[ \tan \theta - \mu_{0} - \frac{u}{V_{0}} + \frac{h}{H_{0}} \right] 
\label{qmd.stat.2} 
\end{equation} 
 
Let $\delta u$ and $\delta h$ be the deviations from the x-uniform regime: 
 
\begin{eqnarray} 
\left\{ \begin{array}{lcl} 
u & = & u_{\infty} + \delta u \\ 
h & = & h_{\infty} + \delta h 
\end{array} \right. 
\end{eqnarray} 
 
Equation~(\ref{qmd.stat.2}) becomes: 
 
\begin{equation} 
\frac {d}{dx} \left[ \alpha  Q  \delta u + \frac{k  g  \cos \theta}{2} \left( 2  h_{\infty} 
 \delta h + \delta h^{2} \right) \right] = g  (h_{\infty} + \delta h)  \cos \theta  \left[ 
\frac{\delta h}{H_{0}} - \frac{\delta u}{V_{0}} \right] 
\label{qmd.stat.3} 
\end{equation} 
 
We now turn all lengths and velocities without dimension by dividing
them respectively by $h_{\infty}$ and by $u_{\infty}$ and define: 
 
\begin{eqnarray} 
\left\{ \begin{array}{lcl} 
\tilde{x} & = & x/h_{\infty} \\ 
\tilde{h} & = & (h-h_{\infty})/h_{\infty} \\ 
\tilde{u} & = & (u-u_{\infty})/u_{\infty} 
\end{array} \right. 
\end{eqnarray} 
 
Mass conservation (equation~(\ref{masse.stat})) implies: 
 
\begin{equation} 
\tilde{h}  \tilde{u} + \tilde{h} + \tilde{u} = 0 
\label{masse.adim} 
\end{equation} 
At this stage let's notice that if $\tilde{h}$ tends to zero, because of equation~(\ref{masse.adim}), 
$\tilde{u}$ also tends to zero (consequence of mass conservation) (eg) the
x-uniform regime is reached. Hence in what follows we focus our 
study on $\tilde{h}$. For its part, equation~(\ref{qmd.stat.3}) becomes: 
\begin{equation} 
\frac{d}{d\tilde{x}}  \left[ \alpha Fr^{2}  \tilde{u} + k  \tilde{h} +
  \frac{1}{2}   k   \tilde{h}^{2} \right] = (1+\tilde{h})  (\frac
{h_{\infty}}{H_0}  \tilde{h} - \frac{u_{\infty}}{V_0}  \tilde{u}) 
\label{qmd.adim} 
\end{equation} 
 
with $Fr$ the Froude number of the x-uniform regime: 
 
\begin{equation} 
Fr = \frac{u_{\infty}}{\sqrt{g  h_{\infty}  \cos \theta}} 
\end{equation} 
 
Substituting, in equation~(\ref{qmd.adim}), $\tilde{u}$ by its expression obtained from 
equation~(\ref{masse.adim}) and introducing the parameter 
$\beta=u_{\infty}  H_{0}/(h_{\infty}  V_{0})$, gives: 
 
\begin{equation} 
\frac{d}{d\tilde{x}}   \left[ k  \tilde{h} + \frac{k}{2}  \tilde{h}^{2}
  - \alpha   Fr^{2}  \frac{\tilde{h}}{\tilde{h}+1} \right] = 
\frac{h_{\infty}}{H_{0}}   \tilde{h}  \left( \tilde{h}+1+\beta \right) 
\end{equation} 

which can be rewritten in: 
 
\begin{equation} 
\left[k  \frac{\tilde{h}+1}{\tilde{h}  (\tilde{h}+1+\beta)} - \alpha  Fr^{2}
  \frac{1}{\tilde{h}  (\tilde{h}+1+\beta)  (\tilde{h}+1)^{2}} \right]   d\tilde{h} = 
\frac{h_{\infty}}{H_{0}}  d\tilde{x} 
\label{qmd.adim.2} 
\end{equation} 
 
A straightforward integration of equation~(\ref{qmd.adim.2}) leads to: 
 
\begin{eqnarray}
\lefteqn{\mbox{\hspace*{-1cm}}\ln \left[ \left| \tilde{h}\right|^{k-\alpha  Fr^{2}}  \left| \tilde{h}+1 \right|^{\alpha  Fr^{2}  
\left( 1-1/\beta^{2} \right) }  \left| \tilde{h}+1+\beta \right|^{k   \beta + \alpha  Fr^{2}/\beta^{2}} 
\right]}\nonumber \\
&&-\alpha  Fr^{2}  \frac{1+\beta}{\beta}  \frac{1}{\tilde{h}+1}
= \left( 1 + \beta \right)  \frac{h_{\infty}}{H_{0}}  \tilde{x} + cst
\end{eqnarray} 
 
with 
 
\begin{eqnarray} 
\left\{ \begin{array}{lcl} 
h_{\infty} & = & \displaystyle \frac{H_{0}  \left( \tan \theta - \mu_{0} \right)}{2}  \left[ -1 
+ \sqrt{1+\varepsilon} \right] \\ 
u_{\infty} & = & \displaystyle \frac{V_{0}  \left( \tan \theta - \mu_{0} \right)}{2}  \left[ 1 
+ \sqrt{1+\varepsilon} \right] \\ 
\varepsilon & = & \displaystyle \frac{4Q}{V_{0}  H_{0}  \left( \tan \theta - \mu_{0} 
\right)^{2}}\\ 
Fr^{2} & = & \displaystyle \frac{u_{\infty}^{2}}{g  h_{\infty}  \cos \theta}\\ 
\beta & = & \displaystyle \frac{u_{\infty} 
  H_{0}}{h_{\infty}  V_{0}} = \frac{\left( 1+\sqrt{ 1
+\varepsilon} \right)^{2}}{\varepsilon} 
\end{array} \right. 
\end{eqnarray}
\newpage

\bibliographystyle{plainnat} 
\bibliography{biblio} 
\end{document}